\documentclass[seceq,preprint]{ptptex}

\usepackage{amsmath}
\usepackage{amssymb}

\notypesetlogo                       

\allowdisplaybreaks

\newcommand{\wt}{\widetilde}
\newcommand{\wh}{\widehat}

\newcommand{\del}{\partial}

\newcommand{\nn}{\nonumber}
\newcommand{\half}{\frac{1}{2}}

\newcommand{\tr}{\mathop{\rm tr}\nolimits}

\newcommand{\cB}{{\mathcal B}}
\newcommand{\cG}{{\mathcal G}}
\newcommand{\cL}{{\mathcal L}}

\newcommand{\cO}{{\mathcal O}}
\newcommand{\cA}{{\mathcal A}}
\newcommand{\cF}{{\mathcal F}}
\newcommand{\cH}{{\mathcal H}}
\newcommand{\bR}{\mathbb{R}}
\newcommand{\bZ}{\mathbb{Z}}

\newcommand{\ba}{{\bf a}}

\newcommand{\Mkk}{M_{\rm KK}}

\newcommand{\gym}{g_{\rm YM}}

\newcommand{\cl}{{\rm cl}}

\newcommand{\Phib}{\overline{\Phi}}

\newcommand{\Ab}{\overline{A}}
\newcommand{\Db}{\overline{D}}

\newcommand{\xt}{\widetilde{x}}

\newcommand{\cAt}{\widetilde{\cA}}
\newcommand{\cFt}{\widetilde{\cF}}
\newcommand{\Ah}{\widehat{A}}
\newcommand{\Fh}{\widehat{F}}

\newcommand{\ds}{\displaystyle}

\newcommand{\Eqref}[1]{Eq.~\eqref{#1}}
\newcommand{\hspc}{\hspace{-2.3mm}}
\renewcommand{\JHEP}{J.\ High Energy Phys.\ }

\title{
{\bf Baryons from Instantons in Holographic QCD}
}

\author{
Hiroyuki \textsc{Hata},$^{1,}$\footnote{
E-mail: {\tt hata@gauge.scphys.kyoto-u.ac.jp}}
\
Tadakatsu \textsc{Sakai},$^{2,}$\footnote{
E-mail: {\tt tsakai@mx.ibaraki.ac.jp}}
\
Shigeki \textsc{Sugimoto}$^{3,}$\footnote{
E-mail: {\tt sugimoto@eken.phys.nagoya-u.ac.jp}}\\
and Shinichiro \textsc{Yamato}$^{1,}$\footnote{
E-mail: {\tt yamato@gauge.scphys.kyoto-u.ac.jp}}
}

\inst{
$^1$Department of Physics, Kyoto University, Kyoto 606-8502, Japan\\
$^2$Department of Physics, Ibaraki University, Mito 310-8512, Japan\\
$^3$Department of Physics, Nagoya University, Nagoya 464-8602, Japan
}

\recdate{
March 6, 2007}

\abst{
We consider aspects of dynamical baryons in a holographic dual of
QCD that is formulated on the basis of a D4/D8-brane configuration.
We construct a soliton solution carrying a unit baryon number and
show that it is obtained as an instanton solution of four-dimensional
Yang-Mills theory with fixed size.
The Chern-Simons term on the flavor D8-branes plays a crucial role
of protecting the instanton from collapsing to zero size.
By quantizing the collective coordinates of the soliton,
we derive the baryon spectra.
Negative-parity baryons as well as baryons with higher
spins and isospins can be obtained in a simple manner.
}

\begin{document}

\maketitle

\section{Introduction}

Since the discovery of the AdS/CFT correspondence
\cite{Maldacena:1997re,Gubser:1998bc,Witten:1998qj}
(For a review, see Ref.~\citen{ADS/CFT}.), it has been recognized that
a gravity description is a promising framework for understanding
non-perturbative aspects of gauge theory. The application of this idea
to realistic models, like QCD, has attracted much attention. (See,
for example, Refs.~\citen{Son:2003et,
Babington:2003vm,Kruczenski:2003uq,Erlich:2005qh,DaRold:2005zs}
for recent progress along this line.)

In Refs.~\citen{SaSu1} and \citen{SaSu2}, a holographic dual of QCD
with $N_f$ massless quarks is constructed using a D4/D8-brane
configuration in type IIA string theory.
It has been argued that the low energy phenomena of QCD, such as
chiral symmetry breaking, can be derived from this model.
The key components of the D4/D8 model are the $G=U(N_f)$
five-dimensional Yang-Mills (YM) and Chern-Simons (CS) theory on a
curved background, both of which originate from the low energy
effective action on the probe D8-branes embedded into the D4
background presented in Ref.~\citen{Witten:D4}.
In this model, the massless pion and an infinite tower of massive
(axial-)vector mesons are interpreted as Kaluza-Klein states
associated with the fifth (or holographic) direction, and the masses
and couplings of the mesons are found to be in good agreement with
experiments. In addition to the mesonic states, dynamical baryons are
also studied in Ref.~\citen{SaSu1}, where it is demonstrated that the
baryon number can be identified with the instanton number of the 5d
YM, and hence it is concluded that baryons can be described by a
soliton with a non-trivial instanton number. (See also
Ref.~\citen{Son:2003et}.) 

In the context of the AdS/CFT correspondence, it has been argued
that baryons are constructed from D-branes wrapped on non-trivial
cycles\cite{
Witten:1998xy,Gross:1998gk,Brandhuber:1998xy,Imamura:1998hf,
Imamura:1998gk,Callan:1998iq,Callan:1999zf}.
In the case of the D4/D8 model,
baryons are identified as D4-branes wrapped
on a non-trivial four-cycle in the D4 background. Such a D4-brane
is realized as a small instanton configuration
in the world-volume gauge theory on the probe D8-brane.
Also, it has been found that the pion effective action obtained from
the 5d YM theory is identically that of the Skyrme model,
in which baryons appear as solitons, called Skyrmions
\cite{Skyrme:1,Skyrme:2,Skyrme:3}.
It can be shown that the baryon number of a Skyrmion,
which is defined as the winding number carried by the pion field,
is equivalent to the instanton number in the 5d YM theory.
In this way, the D4/D8 model connects various descriptions of
baryons. (For further studies of baryons in the AdS/CFT or AdS/QCD,
see Refs.~\citen{Berenstein:2002ke,deTeramond:2004qd,Imamura:2005sw,
Hong:2006ta,Chernicoff:2006yp,NaSuKo}.
Also, closely related works are presented in Refs.~\citen{Son:2003et}
and \citen{Eto:2005cc}.)

The purpose of this paper is to investigate aspects of baryons
described as instantons in the 5d YM-CS theory formulated in the
D4/D8 model. For brevity, we restrict ourselves to the two-flavor
case, $N_f=2$. We first construct a soliton solution of the 5d YM that
carries a unit baryon number.
We show that for $\lambda=\gym^2N_c\gg 1$, which ensures the validity
of the supergravity approximation, the soliton is represented by a
BPST one-instanton solution \cite{BPST} with a fixed size of order
$\lambda^{-1/2}$ located at the origin in the holographic
direction.
Here, the CS term and the $U(1)$ part of the gauge field play an
important role in stabilizing the instanton size.
We next quantize the soliton by formulating a quantum-mechanical
system that governs the collective motion.
A baryon is identified with a quantum state of this system.
Note that this procedure is a natural extension of the old, well-known
idea of Adkins, Nappi and Witten \cite{ANW} in the context of the
Skyrme model \cite{Skyrme:1,Skyrme:2,Skyrme:3}.
In the original work, appearing in Ref.~\citen{ANW}, only the massless
pion is taken into account. Therefore it is natural to extend the
analysis to include the contribution from massive (axial-)vector
mesons.
Such an extension has been studied in Refs.~\citen{Adkins-Nappi,
Igarashi:1985et,Fujiwara:1984pk,Meissner-Zahed}
(See Refs.~\citen{Zahed:1986qz} and \citen{Meissner:1987ge}
for reviews and references therein.)
using phenomenological effective actions including the (axial-)vector
mesons, such as the $\rho$, $\omega$ and $a_1$ mesons.
This paper proposes a new approach for incorporating vector mesons.
This approach utilizes that fact that, in the D4/D8 model, the pion
and an infinite number of the massive (axial-)vector mesons are
unified in a single 5d gauge field with a reasonably simple effective
action.
Thus, it is expected that a thorough study of this model will allow
us to gain some new insight into baryon physics that cannot be
captured by the Skyrme model.

The idea of describing baryons in terms of YM instantons was
previously investigated in Ref.~\citen{AtMa}, in which it is argued
that the pion field configuration corresponding to the Skyrmion
is accurately approximated by integrating the one-instanton solution
along an artificial fifth direction.
Our approach is a manifestation of this idea, although the motivation
is completely different. An interesting point here is that the
introduction of the fifth direction is not just a mathematical trick.
Rather, this direction has a physical interpretation as one of the
spatial directions in the holographic description of QCD.

Unfortunately, because the instanton size is of order
$\lambda^{-1/2}$, it is necessary to incorporate an infinite number of
higher-derivative terms into the 5d YM-CS theory in order to derive
quantitatively precise results concerning baryon physics.
In this paper, we do not attempt to resolve this issue. Instead,
we mainly consider the 5d YM theory with the CS term (although in
Appendix \ref{higher}, we also analyze the non-Abelian DBI action).
For this reason, it may be the case that quantitative comparisons of
our results with experiments, which are made below for several
examples, are of limited physical meaning. However, even if this is
the case, we believe that the qualitative picture of baryon physics
investigated in this paper is rather interesting and can help us to
gain deeper insight into it.
In fact, the baryon spectrum obtained in this paper seems
to capture some characteristics of the baryon spectra observed in
experiments, although the predicted masses are not very close to the
experimental values.

The organization of this paper is as follows.
In \S 2, we formulate the 5d YM-CS system that we treat throughout
this paper.
In \S 3, we show that baryons are described by an instanton solution
whose size is fixed by taking into account the effect of the CS term.
Section 4 is devoted to the construction of the Lagrangian of the
collective motion of the soliton.
Quantization of the Lagrangian is performed in \S 5, where the
correspondence between each quantum state and a baryon is
established. There, we also make a quantitative comparison of our
results with experimental results for several cases.
We end this paper with conclusions in \S 6.
Some technical details are summarized in the Appendices.

\section{The model}
\label{setup}

Our model consists of the following YM-CS theory with gauge group
$U(N_f)$ in a five-dimensional curved background:
\begin{align}
S&=S_{\rm YM}+S_{\rm CS}\ ,\nn\\
S_{\rm YM}&=-\kappa
\int d^4 x dz\,\tr\left[\,
\half\,h(z){\cF}_{\mu\nu}^2+k(z){\cF}_{\mu z}^2
\right]\ ,
\nn\\
S_{\rm CS}&=\frac{N_c}{24\pi^2}
\int_{M^4\times\bR}\omega_5^{U(N_f)}({\cA})\ .
\label{model}
\end{align}
Here, $\mu,\nu=0,1,2,3$ are four-dimensional Lorentz indices, and $z$
is the coordinate of the fifth-dimension. The quantity
${\cA}=\cA_\mu dx^\mu+\cA_z dz$ is the 5-dimensional
$U(N_f)$ gauge field, and $\cF=d\cA+i\cA\wedge\cA$ is
its field strength. The constant $\kappa$ is related to the 't~Hooft
coupling $\lambda$ and the number of colors $N_c$ as
\footnote{
In the early versions of Refs.~\citen{SaSu1} and \citen{SaSu2},
we used $\kappa=\lambda N_c/(108\pi^3)$, which is
due to the misleading factor of 2 appearing in Eq.\ (5.1) of
Ref.~\citen{SaSu1}.}
\begin{equation}
\kappa=\frac{\lambda N_c}{216\pi^3}\equiv a\lambda N_c\ .
\label{kappa}
\end{equation}
The functions $h(z)$ and $k(z)$ are given by
\begin{equation}
h(z)=(1+z^2)^{-1/3}\ ,\quad k(z)=1+z^2\ ,
\label{hk}
\end{equation}
and $\omega_5^{U(N_f)}(\cA)$ is the CS 5-form
for the $U(N_f)$ gauge field defined as
\begin{equation}
\omega_5^{U(N_f)}({\cA})=\tr\left(
\cA \cF^2-\frac{i}{2}\cA^3\cF-\frac{1}{10}\cA^5
\right)\ .
\end{equation}
This theory is obtained as the effective action of $N_f$ probe
D8-branes placed in the D4-brane background studied in
Ref.~\citen{Witten:D4} and is supposed to be an effective theory of
mesons, including an infinite number of (axial-)vector mesons as well
as the massless pion, in four-dimensional QCD with $N_f$ massless
quarks.
In Refs.~\citen{SaSu1} and \citen{SaSu2}, it is argued that much of
the low energy behavior of QCD is reproduced by this simple action.
Here we employ units in which $M_{\rm KK}=1$, where $M_{\rm KK}$ is
the single mass parameter of the model, which specifies the
Kaluza-Klein mass scale. The $M_{\rm KK}$ dependence can easily be
recovered through dimensional analysis.

Note that it is also possible to extend our investigation to cases of
more general functions $h(z)$ and $k(z)$, as in the phenomenological
approach to holographic QCD given in Ref.~\citen{Son:2003et}.
However, in this paper we use the functional forms given in
\Eqref{hk} for definiteness.

It is useful to decompose the $U(N_f)$ gauge field $\cA$
into its $SU(N_f)$ part $A$ and its $U(1)$ part $\wh A$ as
\begin{equation}
\cA=A+\frac{1}{\sqrt{2N_f}}\,\wh A
=A^aT^a+\frac{1}{\sqrt{2N_f}}\,\wh A\ ,
\end{equation}
where $T^a$ ($a=1,2,\cdots,N_f^2-1$) are the generators
for $SU(N_f)$ normalized as
\begin{equation}
\tr(T^aT^b)=\half\delta^{ab}\ .
\end{equation}
The action is then written as
\begin{align}
S_{\rm YM}&=-\kappa
\int d^4 x dz\,\tr\left[\,
\half\,h(z)F_{\mu\nu}^2+k(z)F_{\mu z}^2
\right]
-\frac{\kappa}{2}\int d^4 x dz\left[\,
\half\,h(z)\Fh_{\mu\nu}^2+k(z)\Fh_{\mu z}^2
\right]\ ,
\\
S_{\rm{CS}}&=\frac{N_c}{24\pi^2}\int\biggl[\,
\omega_5^{SU(N_f)}(A)
+\frac{3}{\sqrt{2N_f}}\Ah\tr F^2
+\frac{1}{2\sqrt{2N_f}}\Ah\,\Fh^2
\nn\\
&\qquad
+\frac{1}{\sqrt{2N_f}}\,
d\left(\Ah\tr\left(2FA-\frac{i}{2}A^3\right)\right)
\biggr] \ .
\label{CS:u_su}
\end{align}

As mentioned above, we consider only the $N_f=2$ case in the present
paper. In this case, $\omega_5^{SU(2)}(A)$ vanishes, and the CS term
reduces to 
\begin{align}
S_{\rm{CS}}
&=\frac{N_c}{24\pi^2}
\int\biggl[\,
\frac{3}{2}\Ah\tr F^2
+\frac{1}{4}\Ah\,\Fh^2
+\mbox{(total derivatives)}\biggr]
\nn\\
&=\frac{N_c}{24\pi^2}
\epsilon_{MNPQ}\int d^4 x dz\biggl[\,
\frac{3}{8}\Ah_0\tr( F_{MN}F_{PQ})
-\frac{3}{2}\Ah_M\tr( \del_0A_NF_{PQ})
\nn\\
&\quad
+\frac{3}{4}\Fh_{MN}\tr( A_0F_{PQ})
+\frac{1}{16}\Ah_0\Fh_{MN}\Fh_{PQ}
-\frac{1}{4}\Ah_M\Fh_{0N}\Fh_{PQ}
+\mbox{(total derivatives)}\biggr] \ ,
\label{CS}
\end{align}
with $M,N=1,2,3,z$ and $\epsilon_{123z}=+1$.

\section{Classical solution}
\label{solution}

\subsection{Soliton solutions for $S_{\rm YM}$}
\label{YM}

In our model, $\lambda$ is assumed to be large, and we employ the
$1/\lambda$ expansion. Since $S_{\rm YM}\sim\cO(\lambda^1)$ and
$S_{\rm CS}\sim\cO(\lambda^0)$, it is expected that the leading
contribution to the soliton mass comes from $S_{\rm YM}$.
Let us first consider the system without the CS term.
In this case, the $U(1)$ part, $\Ah$, of the gauge field is
decoupled from the $SU(2)$ part, and thus it is consistent to
set $\Ah=0$. We are interested in the minimal energy
static configuration carrying a unit baryon number, $N_B=1$, where the
baryon number $N_B$ is equal to the instanton number and is given by
\begin{equation}
N_B=\frac{1}{32\pi^2}\int\! d^3xdz\,\epsilon_{MNPQ}
\tr(F_{MN}F_{PQ})\ .
\end{equation}

If the five-dimensional space-time were flat and the functions $h(z)$
and $k(z)$ were trivial (i.e.\ $h(z)=k(z)=1$), the solution would be
given by the BPST instanton solution \cite{BPST} of arbitrary size
$\rho$ and position in the four-dimensional space parameterized by
$x^M$ ($M=1,2,3,z$).
However, in the present case with \Eqref{hk}, it can be shown that
the minimal energy configuration is given by a small instanton with
infinitesimal size, $\rho\to 0$.

To illustrate this fact, we first examine the $\rho$ dependence
of the energy calculated by inserting the BPST instanton
configuration as a trial configuration.
The BPST instanton configuration is given by
\begin{equation}
A_M(x)=-if(\xi)\,g\del_Mg^{-1} \ ,
\label{eq:BPST}
\end{equation}
where
\begin{align}
f(\xi)&=\frac{\xi^2}{\xi^2+\rho^2} \ ,\quad
\xi=\sqrt{(\vec{x}-\vec X)^2+(z-Z)^2} \ ,
\label{eq:f}
\\
g(x)&=\frac{(z-Z)-i(\vec{x}-\vec X)\cdot\vec{\tau}}{\xi} \ ,
\label{eq:g}
\end{align}
and its field strengths are
\begin{equation}
F_{ij}=\frac{2\rho^2}{(\xi^2+\rho^2)^2}\epsilon_{ija}\tau^a \ ,
\quad
F_{zj}=\frac{2\rho^2}{(\xi^2+\rho^2)^2}\tau_j \ .
\end{equation}
Here  $\vec\tau=(\tau^1,\tau^2,\tau^3)$ are the Pauli matrices,
and we have $\vec x=(x^1,x^2,x^3)$ and $a,i,j=1,2,3$. 
The constants $(\vec X,Z)$ and $\rho$ denote the position
and the size of the instanton, respectively.
This is the one-instanton solution for the $SU(2)$ Yang-Mills theory
in flat four-dimensional space.
Assuming $\vec X=0$ and $Z=0$ for simplicity,
the energy of this configuration is calculated as
\begin{align}
E(\rho)&=\kappa\int d^3xdz\,\tr\left[
\frac{1}{2}h(z)F_{ij}^2+k(z)F_{iz}^2
\right]\nn\\
&=3\pi^2\kappa\rho^4\int dz\,(z^2+\rho^2)^{-5/2}(h(z)+k(z))\nn\\
&=3\pi^2\kappa
\left[
\frac{\sqrt{\pi}\,\Gamma(7/3)}{\Gamma(17/6)}
\,F\left(\frac{1}{3},\frac{1}{2},\frac{17}{6};1-\rho^2\right)
+\frac{4}{3}+\frac{2}{3}\rho^2
\right]\ .
\label{bpstenergy}
\end{align}
It can be shown that $E(\rho)$ is a monotonically increasing function
of $\rho$ whose minimal value is $E(\rho=0)=8\pi^2\kappa$.

It is also possible to show that the minimal value of the energy
$E=8\pi^2\kappa$ found above is actually the absolute minimum in the
sector with a unit instanton number. In fact, the $SU(2)$ part of
the YM action has the following bound for any static configuration:
\begin{align}
\kappa \int d^3x dz\tr
\left[\frac{1}{2}h(z)F^2_{ij}+k(z)F^2_{iz}\right]
&\geq \frac{\kappa}{2}\int d^3x dz
\sqrt{h(z)k(z)}\left|\epsilon ^{ijk}F^a_{jk}F^a_{iz}\right|
\nn\\
&\geq 8\pi^2\kappa |N_B|\ .
\label{inequality}
\end{align}
Here we have used the relation $h(z)k(z)\ge h(0)k(0)=1$.
The lower bound of \eqref{inequality} is realized only in the case of
a (anti-)self-dual instanton with an infinitesimal size located at
$z=0$.

It is interesting that the minimal value $8\pi^2\kappa$ is equal to
the baryon mass obtained in Ref.~\citen{SaSu1} from the mass of a
D4-brane wrapped around an $S^4$ that surrounds the color D4-branes.
This fact suggests that the soliton mass $8\pi^2\kappa$ is not
modified even if we include higher derivative terms in the DBI
action, because the wrapped D4-brane can be regarded as a small
instanton on the probe D8-branes \cite{Witten:1995gx,Douglas:1995bn}.
More evidence supporting this conjecture is given in Appendix
\ref{higher}.

\subsection{Contribution from $S_{\rm CS}$}
\label{CSterm}

Let us next consider the contribution from the CS term, \eqref{CS}.
It is important to note that this term includes a term of the form
\begin{equation}
\epsilon_{MNPQ}\int d^4xdz\,\Ah_0 \tr( F_{MN}F_{PQ}) \ .
\label{AFF}
\end{equation}
This shows that the instanton configuration induces an electric
charge coupled to the $U(1)$ gauge field $\Ah$.
As is well known from the theory of electrodynamics, the energy
possessed by the electric field of a point charge diverges.
In the 1+4 dimensional case, the energy behaves as $E\sim\rho^{-2}$
for a charged particle of radius $\rho$.
Then, taking this contribution into account, it follows that the
minimal energy configuration representing a baryon must have a
finite size. This reasoning is analogous to that used to argue the
stability of a Skyrmion via the $\omega$ meson presented in
Ref.~\citen{Adkins-Nappi}.

In fact, as we show below, the classical solution at leading
order in the $1/\lambda$ expansion is given by a BPST instanton in the
flat space whose size $\rho$ is of order $\lambda^{-1/2}$.
For this reason, in order to carry out a systematic $1/\lambda$
expansion, it is convenient to rescale the coordinates $x^M$ as well
as the $U(2)$ gauge field $\cA_M$ as
\begin{align}
&\xt^M=\lambda^{+1/2}x^M \ ,\quad \xt^0=x^0 \ ,
\nn\\
&\cAt_0(t,\xt)=\cA_0(t,x) \ ,\quad
\cAt_M(t,\xt)=\lambda^{-1/2}\cA_M(t,x) \ ,
\nn\\
&\cFt_{MN}(t,\xt)=\lambda^{-1}\cF_{MN}(t,x) \ ,\quad
\cFt_{0M}(t,\xt)=\lambda^{-1/2}\cF_{0M}(t,x) \ ,
\label{rescaling}
\end{align}
and regard the quantities with tildes as being $\cO(\lambda^0)$.
Hereafter, we omit the tilde for simplicity.
We then find that for $\lambda\gg 1$, the YM part becomes
\begin{align}
S_{{\rm YM}}=&-aN_c\int d^4 x dz \,\tr\left[\,
\frac{\lambda}{2}\,F_{MN}^2+\left(
-\frac{z^2}{6} F_{ij}^2+z^2 F_{iz}^2- F_{0M}^2
\right)+\cO(\lambda^{-1})
\right]
\nn\\
&-\frac{aN_c}{2}\int d^4 x dz \,
\left[\,\frac{\lambda}{2}
\, \Fh_{MN}^2+
\left(
-\frac{z^2}{6} \Fh_{ij}^2+z^2 \Fh_{iz}^2-\Fh_{0M}^2
\right)+\cO(\lambda^{-1})\right]\ ,
\label{DBI}
\end{align}
with $i,j=1,2,3$, while the CS term takes the same form as that given
in \Eqref{CS}. Here we have used \Eqref{kappa}.
The equations of motion for the $SU(2)$ part read
\begin{align}
&D_M F_{0M}+\frac{1}{64\pi^2 a}
\epsilon_{MNPQ}\Fh_{MN}F_{PQ}
+\cO(\lambda^{-1})=0 \ ,
\label{Gausslaw}\\
&D_NF_{MN}+\cO(\lambda^{-1})=0 \ .
\label{eom:su}
\end{align}
Also, the equations of motion for the $U(1)$ part are
\begin{align}
&\del_M\Fh_{0M}+\frac{1}{64\pi^2 a}
\epsilon_{MNPQ}\left\{
\tr(F_{MN}F_{PQ})+\half\Fh_{MN}\Fh_{PQ}
\right\}+\cO(\lambda^{-1})=0 \ ,
\label{Gauss:u1}\\
&\del_{N}\Fh_{MN}+\cO(\lambda^{-1})=0 \ .
\label{eom:u1}
\end{align}

Now we solve the equations of motion \eqref{Gausslaw}--\eqref{eom:u1}
in order to derive a static soliton solution corresponding to a
baryon. First, let us consider \Eqref{eom:su}.
In this paper we expand the action about the baryon solution and keep
only the terms of orders $\lambda^1$ and $\lambda^0$.
For this purpose, we have only to solve the equation $D_NF_{MN}=0$ on
flat space while ignoring the $\cO(\lambda^{-1})$ term in
\Eqref{eom:su}, because the correction to the solution from the
$\cO(\lambda^{-1})$ term in \Eqref{eom:su} gives only an
$\cO(\lambda^{-1})$ correction to the action. Therefore, a solution
that carries a unit baryon number is given by the BPST instanton
solution \eqref{eq:BPST}. Here, the parameters $(\vec X,Z)$ and $\rho$
are also rescaled as in \Eqref{rescaling}.

For the $U(1)$ part, the finite energy solution of the Maxwell
equation \eqref{eom:u1} is given by $\wh F_{MN}=0$, which yields the
trivial solution $\Ah_M=0$, up to a gauge transformation.
Then the Gauss's law equation \eqref{Gausslaw} is reduced to
\begin{equation}
D_M^2 A_0=0\ ,
\end{equation}
whose solution is given in terms of a linear combination of the
functions $\Phi_a$ given in \Eqref{phia} of Appendix
\ref{metric:inst}. We are interested in the solution that vanishes at
infinity, and it is given by $A_0=0$.

We are thus left with \Eqref{Gauss:u1} for $\Ah_0$:
\begin{equation}
\del_M^2\Ah_0+\frac{3}{\pi^2a}
\frac{\rho^4}{(\xi^2+\rho^2)^4}=0 \ .
\end{equation}
This equation can easily be solved, and the regular solution that
vanishes at infinity is given by
\begin{equation}
\Ah_0=\frac{1}{8\pi^2a}\frac{1}{\xi^2}
\left[1-\frac{\rho^4}{(\rho^2+\xi^2)^2}
\right] \ .
\label{A0hat}
\end{equation}
Note that we could add a constant term to \Eqref{A0hat} if we allow
$\Ah_0$ that are non-vanishing at infinity.
The physical interpretation of this constant term is that it is the
chemical potential $\mu$ associated with the baryon number,\footnote{
See Refs.~\citen{Kim-Sin-Zahed,Tanii-Horigome,Parnachev:2006ev}
for recent developments concerning the D4/D8 model with a chemical
potential. 
}
since \Eqref{AFF} induces the $\mu N_B$ term in the action.

Now we have obtained the configurations \eqref{eq:BPST} and
\eqref{A0hat}, together with $A_0=\Ah_M=0$, which solves the
leading-order equations of motion, \eqref{Gausslaw}--\eqref{eom:u1}.
Although this solution is sufficient for calculating the
$\cO(\lambda^1)$ and $\cO(\lambda^0)$ terms of the energy, as
mentioned below \Eqref{eom:u1}, the resultant energy depends on $\rho$
and $Z$, which have not yet been fixed.
In fact, the soliton mass $M$ is obtained by evaluating the action
on shell, $S=-\int dt M$:
\begin{align}
M&=8\pi^2\kappa+
\kappa\lambda^{-1}\int d^3xdz
\left[-\frac{z^2}{6}\tr (F_{ij})^2+z^2\tr (F_{iz})^2\right]
\nn\\
&\qquad
-\half\kappa\lambda^{-1}\int d^3xdz\left[
(\del_M\Ah_0)^2
+\frac{1}{32\pi^2a}\Ah_0\,\epsilon_{MNPQ}\tr(F_{MN}F_{PQ})\right]
+\cO(\lambda^{-1})\nn\\
&=8\pi^2\kappa\left[
1+\lambda^{-1}\left(\frac{\rho^2}{6}
+\frac{1}{320\pi^4a^2}
\frac{1}{\rho^2}+\frac{Z^2}{3}\right)
+\cO(\lambda^{-2})\right] \ .
\label{smass}
\end{align}
The values of $\rho$ and $Z$ for the solution should be determined by
minimizing $M$, which is equivalent to solving the sub-leading part of
the equations of motion, \eqref{eom:su} and \eqref{eom:u1}, projected
on to the space of the deformations of the solution in the $\rho$ and
$Z$ directions.

It is worth emphasizing that the term in \Eqref{smass} proportional to
$\rho^{-2}$ results from the Coulomb interaction
$\Ah_0\,\epsilon_{MNPQ}\tr(F_{MN}F_{PQ})$ in the CS term, while
the $\rho^2$ and $Z^2$ terms are due to the warped geometry employed
here. Without the Coulomb interaction, the soliton mass is
minimized by the instanton with infinitesimal size, i.e.\ $\rho\to 0$,
located at the origin, $Z=0$, as we saw in \S \ref{YM}. However, with
the Coulomb interaction, the instanton is stabilized at a finite
size $\rho$ given by 
\begin{equation}
\rho^2=\frac{1}{8\pi^2 a}\sqrt{\frac{6}{5}} \ .
\label{rhomin}
\end{equation}
Going back to the original variable [see \Eqref{rescaling}],
$\rho^2$ is rescaled as $\rho^2\to\lambda^{+1}\rho^2$, ensuring that
the soliton is given by an instanton with size of order
$\lambda^{-1/2}$, as mentioned above. Then, inserting \Eqref{rhomin}
into \Eqref{smass}, the mass of the soliton becomes
\begin{align}
M&\simeq8\pi^2\kappa
+\sqrt{\frac{2}{15}}N_c \ .
\label{Mcl}
\end{align}

We conclude this section with a few remarks on higher-order derivative
terms. The action \eqref{model} is obtained by omitting the higher
derivative terms from the D-brane effective action.
This corresponds to keeping only the leading-order terms in the
$1/\lambda$ expansion. However, in our case, because the size of the
soliton solution is small, the derivative of the gauge field is
enhanced and may become important in the analysis.
Actually, we have seen that the size of the soliton solution is of
order $\lambda^{-1/2}$, which in turn implies that an infinite
number of higher-derivative terms involved in the D-brane effective
action are of the same order in the $1/\lambda$ expansion.
To see this, recall that each derivative and gauge field is
accompanied by the string length $l_s=\sqrt{\alpha'}$ in the DBI
action, for example, $l_s\del_M$, $l_s A_M$ and $\alpha' F_{MN}$.
As explained in Ref.~\citen{SaSu2}, $\alpha'$ can be regarded as a
parameter of order $\lambda^{-1}$. Therefore, after the rescaling of
\Eqref{rescaling}, $l_s\del_M$ and $l_s A_M$ become $\cO(\lambda^0)$
in the rescaled variables, and hence the higher-order derivative terms
can appear at the same order. Such terms may also contribute to the
equations of motion, \eqref{Gausslaw}--\eqref{eom:u1}, and the soliton
mass \eqref{smass}.
On the other hand, there are some arguments indicating that, in the
case of D-branes in a flat space-time, neither the BPST instanton
solution nor its energy is modified, even if all the higher
derivative corrections are taken into account
\cite{Hashimoto-Terashima,9711097,9801127,9804180}.
In Appendix \ref{higher}, we investigate the non-Abelian DBI action
and obtain some evidence that the analysis based on the Yang-Mills
action given in \Eqref{model} is not modified. It is important to
carry out a more systematic analysis in order to make precise 
quantitative predictions. We leave this task for a future study.

\section{Lagrangian of the collective modes}
\label{L}

The moduli space of the one-instanton solution for the $SU(2)$
Yang-Mills equation \eqref{eom:su}, ignoring the $\cO(\lambda^{-1})$
terms, is given by
\begin{equation}
{\cal M}=\bR^4\times \bR^4/\bZ_2\ .
\label{moduli}
\end{equation}
The first $\bR^4$ here corresponds to the position of the instanton
parameterized by $(\vec X,Z)$, and $\bR^4/\bZ_2$ consists of the size
$\rho$ and the $SU(2)$ orientation of the instanton.
(See, for example, Ref.~\citen{instanton:review} for a review.)
Let us parameterize $\bR^4/\bZ_2$ by $y_I$ ($I=1,2,3,4$), which are
transformed as $y_I\to -y_I$ under $\bZ_2$.
The size of the instanton corresponds to the radial coordinate,
$\rho=\sqrt{y_1^2+\cdots+y_4^2}$, and the $SU(2)$ orientation
is parameterized by $a_I\equiv y_I/\rho$, with
the constraint $\sum_{I=1}^4 a_I^2=1$.

To analyze slowly moving solitons, we adopt the moduli space
approximation method \cite{Gervais:1974dc,Manton}. With this method,
we treat the collective coordinates $(\vec X,Z,y_I)$ as
time-dependent variables and consider a quantum mechanical description
of a particle in the moduli space ${\cal M}$.
The situation here is analogous to that of monopoles \cite{Manton}
(See, for example,
Refs.~\citen{Harvey,instanton:review,Weinberg:2006rq} for a review.)
and also that of the Skyrmions \cite{ANW}.
In the present case, the size $\rho$ and the position $Z$ in the
$z$-direction are not genuine collective coordinates, because of the
$\rho$ and $Z$ dependent terms in the energy \eqref{smass}, which
arises from the non-trivial warp factors $h(z)$ and $k(z)$.
As seen at the end of this section,
the excitations associated with $\rho$ and $Z$ are much lighter
than those associated with the other massive modes around the
instanton for large $\lambda$. For this reason, we treat $\rho$ and
$Z$ as collective coordinates, along with $(\vec X,a_I)$.

Now we calculate the effective Lagrangian of these collective modes,
presenting the derivation of \Eqref{moduli} for completeness.
We work in the $A_0=0$ gauge, which should be accompanied by the
Gauss's law constraint \eqref{Gausslaw}.
Also, \Eqref{Gauss:u1} gives a constraint for obtaining $\Ah_0$ and
singles out the physical degrees of freedom.

The basic idea employed in this calculation is to approximate the
slowly moving soliton by the static classical solution, with the
constant moduli $X^\alpha=(\vec X,Z,y_I)$ promoted to the
time-dependent collective coordinates $X^\alpha(t)$.
Thus, the $SU(2)$ gauge field is assumed to be of the form
\begin{equation}
A_M(t,x)=
VA_M^{\rm cl}(x;X^\alpha(t))V^{-1}
-i\,V\del_M V^{-1} \ .
\label{def:coll}
\end{equation}
Here, $A_M^{\rm cl}(x;X^\alpha(t))$ is the instanton solution
\eqref{eq:BPST} with time-dependent collective coordinates
$\rho(t)$, $\vec X(t)$ and $Z(t)$. The quantity $V=V(t,x)$ is an
element of $SU(2)$ that is necessary for imposing the Gauss's law
constraint \eqref{Gausslaw} for \Eqref{def:coll}.
It also specifies the $SU(2)$ orientation and hence includes
the collective coordinates $a_I(t)$.
To see this, we first note that
\begin{equation}
F_{MN}=VF_{MN}^{\cl}V^{-1} \ ,\quad
F_{0M}=V\left(
\dot X^\alpha \del_\alpha{A}_{M}^{\rm cl}
-D_M^{\cl}\Phi\right)V^{-1} \ ,
\label{F0M}
\end{equation}
where $\del_\alpha=\del/\del X^\alpha$, the dot denotes the time
derivative $\del_0$, $D_M^{\rm cl}$ is the covariant derivative 
with the gauge field $A_M^{\rm cl}(x;X^\alpha(t))$, and we have
\begin{equation}
\Phi\equiv -i V^{-1}\dot{V}\ .
\label{VV}
\end{equation}
For a given $\Phi$, $V$ can be obtained as
\begin{equation}
V^{-1}={\rm P}
\exp\left(-i\int^t dt'\Phi(t',x)\right)\ .
\end{equation}
It then follows that \Eqref{Gausslaw} becomes
\begin{equation}
D_M^{\rm cl}\left(\dot X^N
\frac{\del}{\del X^N} A_M^{\rm cl}
+\dot\rho \frac{\del}{\del \rho} A_M^{\rm cl}
-D_M^{\rm cl}\Phi\right)=0\ ,
\label{Gausslaw2}
\end{equation}
where $X^N=(\vec X,Z)$, and we have used $\Fh_{MN}^{\rm cl}=0$.
As shown in Appendix \ref{metric:inst}, this equation is solved by
choosing
\begin{equation}
\Phi(t,x)=-\dot X^N(t) A_N^{\rm cl}(x)+
\chi^a(t)\Phi_a(x)\ ,
\end{equation}
where $\Phi_a$ ($a=1,2,3$) are the solutions of
$D_M^{\rm cl}D_M^{\rm cl}\Phi_a=0$ given in \Eqref{phia}, and $\chi^a$
($a=1,2,3$) are related to the collective coordinates $a_I$ as
\begin{equation}
\chi^a=-i\tr( \tau^a \ba^{-1}\dot \ba)
=2(a_4\dot a_a-\dot a_4 a_a+\epsilon_{abc}a_b\dot a_c)\ ,
\end{equation}
with
\begin{equation}
\ba\equiv a_4+i a_a\tau^a\ \in SU(2)\ .
\label{ba}
\end{equation}
Then, $F_{0M}$ in \Eqref{F0M} can be expressed as
\begin{equation}
F_{0M}=V\left(
\dot X^NF_{MN}^{\rm cl}
+\dot\rho\frac{\del}{\del\rho}A_M^{\rm cl}
-\chi^a D_M^{\rm cl}\Phi_a
\right)V^{-1}\ ,
\end{equation}
where we have used
$(\del/\del X^N)A_M^{\rm cl}=-\del_N A_M^{\rm cl}$.

It is also necessary to impose the condition represented by
\Eqref{Gauss:u1}, which reads
\begin{equation}
-\del_M^2\Ah_0+\frac{1}{64\pi^2 a}\epsilon_{MNPQ}
\tr(F_{MN}^{\rm cl}F_{PQ}^{\rm cl})=0 \ .
\end{equation}
This shows that $\Ah_0$ is again given by \Eqref{A0hat},
except that in the present case, all the instanton moduli are
time dependent.

Inserting into the action the above soliton configuration with
time-dependent collective coordinates, we obtain the quantum
mechanical system
\begin{align}
L&=\frac{m_X}{2}\, g_{\alpha\beta}\dot X^\alpha\dot X^\beta-
U(X^\alpha)+\cO(\lambda^{-1})\ ,
\label{eq:Lqms}
\end{align}
where $m_X\equiv 8\pi^2\kappa\lambda^{-1}=8\pi^2 aN_c$ and
$g_{\alpha\beta}$ is the metric for the instanton moduli space
\eqref{moduli}, given by
\begin{align}
ds^2&=g_{\alpha\beta}\,dX^\alpha dX^\beta\nn\\
&=d\vec X^2+dZ^2
+2(d\rho^2+\rho^2 da_I^2)\nn\\
&=d\vec X^2+dZ^2
+2\, dy_I^2\ .
\label{metric}
\end{align}
(See Appendix \ref{metric:inst} for more details.)
The potential $U(X^\alpha)$ is given by \Eqref{smass},
\begin{equation}
U(X^\alpha)=U(\rho,Z)=M_0+m_X\left(\frac{\rho^2}{6}
+\frac{1}{320\pi^4 a^2}\frac{1}{\rho^2}+\frac{Z^2}{3}\right)\ ,
\label{UX}
\end{equation}
with $M_0=8\pi^2\kappa$.
The Lagrangian \eqref{eq:Lqms} can also be written as
\begin{align}
L&=L_X+L_Z+L_y+\cO(\lambda^{-1})\ ,
\label{Lag}
\\
L_X&=-M_0+\frac{m_X}{2}\dot{\vec X}^2\ ,\nn\\
L_Z&=\frac{m_Z}{2}\dot{Z}^2-\frac{m_Z\omega_Z^2}{2}Z^2\ ,\nn\\
L_y&=\frac{m_y}{2}\dot{y}_I^2-\frac{m_y\omega_\rho^2}{2}\rho^2
-\frac{Q}{\rho^2}
=\frac{m_y}{2}\left(\dot\rho^2+\rho^2\dot a_I^2\right)
-\frac{m_y\omega_\rho^2}{2}\rho^2
-\frac{Q}{\rho^2}\ ,
\label{Lagy}
\end{align}
where
\begin{align}
&M_0=8\pi^2\kappa \ ,\quad
m_X=m_Z=m_y/2=8\pi^2\kappa\lambda^{-1}=8\pi^2 a N_c\ ,\nn\\
&\omega_Z^2=\frac{2}{3} \ ,\quad\omega_\rho^2=\frac{1}{6} \ ,
\quad Q=\frac{N_c^2}{5m_X}=
\frac{N_c}{40\pi^2 a} \ .
\end{align}

A few comments are in order. First, if we write the above Lagrangian
in terms of the original variable before the rescaling described in
\Eqref{rescaling}, $m_X$ is replaced with $M_0$, and then the
$\dot{\vec X}^2$ term becomes the usual kinetic term for a particle of
mass $M_0$. Second, note that the Lagrangian for $a_I$ is the same as
that in the case of a Skyrmion \cite{ANW} with a moment of inertia
$m_y\rho^2/4$, although this moment of inertia depends on the
coordinate $\rho$, which is promoted to an operator upon
quantization. Third, as mentioned above, $\rho$ and $Z$ are not the
collective modes in the usual sense, since they have the non-trivial
potential \eqref{UX}. The reason that we focus only on $\rho$ and $Z$
among the infinitely many massive fluctuations about the instanton is
the following. Because the Lagrangian for $\rho$ and $Z$ in
\Eqref{Lag} is of order $\lambda^0$, the energy induced by the
excitation of these modes is also of order $\lambda^0$.
On the other hand, the other massive fluctuations are all massive,
even for a flat background, and hence the mass terms come from the
$\cO(\lambda)$ term in \Eqref{DBI}. This implies that their
frequencies are of order $\lambda^{1/2}$.
Therefore, the excitations of these modes are much heavier than the
excitations of $Z$ and $\rho$ for $\lambda\gg 1$.

\section{Quantization}

In this section, we quantize the system \eqref{Lag} in order to derive
the spectra of baryons. The Hamiltonian for a baryon placed at $\vec
X=0$ is
\begin{equation}
H=M_0+H_y+H_Z\ ,
\end{equation}
where
\begin{align}
H_y&= -\frac{1}{2m_y}\sum_{I=1}^4 \frac{\del^2}{\del y_I^2}+
\half m_y\omega_\rho^2 \rho^2
+\frac{Q}{\rho^2}\ ,
\label{Hy}\\
H_Z&= -\frac{1}{2m_Z}\del_Z^2+ \half m_Z\omega_Z^2 Z^2\ .
\label{HZ}
\end{align}

As argued in Appendix \ref{metric:inst}, a point $a_I$ in $S^3$
and its antipodal point, $-a_I$, are to be identified in the instanton
moduli space. This implies that the wave function of the system must
satisfy the condition
\begin{equation}
\psi(a_I)=\pm \psi(-a_I)\ .
\end{equation}
Following Ref.~\citen{ANW} (see also Ref.~\citen{FR}), we impose the
anti-periodic boundary condition $\psi(a_I)=-\psi(-a_I)$, since we are
interested in fermionic states.

\subsection{Solution to the Schr\"odinger equation}

As a warm-up, let us first consider $H_y$ with $Q=0$.
Then, the system is reduced to the 4-dimensional harmonic oscillator:
\begin{equation}
H_y|_{Q=0}=\sum_{I=1}^4 \left(
-\frac{1}{2m_y}\frac{\del^2}{\del y_I^2}+
\half m_y\omega_\rho^2 y_I^2\right)\ .
\end{equation}
We know that the energy eigenvalues of this system are given by
\begin{equation}
E_y|_{Q=0}=\omega_\rho(N+2)\ ,
\label{En}
\end{equation}
with
\begin{equation}
N=n_1+n_2+n_3+n_4\ ,
\end{equation}
where $n_I=0,1,2,\cdots$ ($I=1,2,3,4$).
The degeneracy of the states with a given $N$ is
\begin{equation}
d_N=\frac{1}{6}(N+3)(N+2)(N+1)\ .
\label{dn}
\end{equation}

Next, we solve this problem using polar coordinates.
The Hamiltonian is then written
\begin{equation}
H_y|_{Q=0}=
-\frac{1}{2m_y}\left(
\frac{1}{\rho^3}\del_\rho (\rho^3\del_\rho)
+\frac{1}{\rho^2}\nabla^2_{S^3}\right)
+\half m_y\omega_\rho^2 \rho^2\ ,
\label{Hrho}
\end{equation}
where $\nabla^2_{S^3}$ is the Laplacian for a unit $S^3$.
It is known that the scalar spherical harmonics for $S^3$ are given by
\begin{equation}
T^{(l)}(a_I)=C_{I_1\cdots I_l}\,a_{I_1}\cdots a_{I_l}\ ,
\end{equation}
where $C_{I_1\cdots I_l}$ is a traceless symmetric tensor of rank
$l$. They satisfy
\begin{equation}
\nabla^2_{S^3}T^{(l)}=-l(l+2)T^{(l)}\ ,
\end{equation}
and the degeneracy is $(l+1)^2$.
Under the isomorphism $SO(4)\simeq (SU(2)\times SU(2))/\bZ_2$,
the rank $l$ traceless symmetric tensor representation of $SO(4)$
corresponds to the $(S_{l/2},S_{l/2})$ representation of
$(SU(2)\times SU(2))/\bZ_2$. Here, $S_{l/2}$ denotes the spin $l/2$
representation of $SU(2)$, and its rank is $\dim S_{l/2}= l+1$.
Writing the eigenfunctions of the Hamiltonian as
\begin{equation}
\psi(y_I)=R(\rho)\, T^{(l)}(a_I) \ ,
\label{polar}
\end{equation}
$R(\rho)$ is found to satisfy
\begin{equation}
\cH_l R(\rho)=E_y|_{Q=0}\, R(\rho)\ ,
\label{HlE}
\end{equation}
with
\begin{equation}
\cH_l\equiv -\frac{1}{2m_y}\left(
\frac{1}{\rho^3}\del_\rho (\rho^3\del_\rho)
-\frac{l(l+2)}{\rho^2}\right)+
\half m_y\omega_\rho^2 \rho^2\ .
\label{Hl}
\end{equation}
The eigenvalue equation \eqref{HlE} for $R(\rho)$ is reduced by
substituting the form
\begin{equation}
R(\rho)= e^{-\frac{m_y\omega_\rho}{2}\rho^2}\,\rho^l\,
v(m_y\omega_\rho\rho^2) \ .
\end{equation}
This yields the confluent hypergeometric differential equation for
$v(z)$,
\begin{equation}
\left\{z\del_z^2 +(l+2-z)\del_z+\frac12\left(
\frac{E_y|_{Q=0}}{\omega_\rho}-l-2\right)\right\}v(z)=0 \ .
\label{eq:CHGDEf}
\end{equation}
A normalizable regular solution to \Eqref{eq:CHGDEf} exists only
when
$(1/2)\left(E_y|_{Q=0}/\omega_\rho-l-2\right)=n=
0,1,2,\cdots$, and it is given by
\begin{equation}
v(z)=F(-n,l+2;z) \ ,
\end{equation}
where $F(\alpha,\gamma;z)$ is the confluent hypergeometric function
defined by
\begin{equation}
F(\alpha,\gamma;z)\equiv\sum_{k=0}^\infty
\frac{(\alpha)_k}{(\gamma)_k}\frac{z^k}{k!} \ ,
\end{equation}
with $(\alpha)_k\equiv \alpha(\alpha+1)\cdots(\alpha+k-1)$.
Note that $F(-n,\gamma;z)$ is a polynomial of degree $n$.
The corresponding energy eigenvalue is
\begin{equation}
E_y|_{Q=0}=\omega_\rho(l+2n+2)\ ,
\end{equation}
which coincides with \Eqref{En}. It is easy to see that the
degeneracy \eqref{dn} is reproduced by summing $(l+1)^2$ with $l=N-2n$
over $n=0,1,\cdots,[N/2]$.

Now we turn back to the Hamiltonian \eqref{Hy} with $Q>0$.
Using polar coordinates, it is written
\begin{equation}
H_y= -\frac{1}{2m_y}\left(
\frac{1}{\rho^3}\del_\rho (\rho^3\del_\rho)
+\frac{1}{\rho^2}(\nabla^2_{S^3}-2m_y Q)\right)+
\half m_y\omega_\rho^2 \rho^2\ .
\end{equation}
Again, the wave function can be written as \Eqref{polar},
and $R(\rho)$ should satisfy
\begin{equation}
\cH_{\wt l}R(\rho)=E_yR(\rho)\ ,
\end{equation}
where $\cH_{\wt l}$ is now given by $\cH_l$ (\Eqref{Hl}), with
$l$ replaced by $\wt{l}$, defined as
\begin{equation}
\wt l\equiv -1+\sqrt{(l+1)^2+2m_y Q}\ ,
\end{equation}
which satisfies
\begin{equation}
\wt l(\wt l+2)=l(l+2)+2m_y Q\ .
\end{equation}
Therefore, the eigenfunctions and the energy eigenvalues are obtained
by simply replacing $l$ with $\wt l$ in the previous results for
$Q=0$, and thus the energy spectrum becomes
\begin{align}
E_y&=\omega_\rho(\wt l+2n_\rho+2)
\nn\\
&=\sqrt{\frac{(l+1)^2}{6}
+\frac{2}{15}N_c^2}+\frac{2n_\rho+1}{\sqrt{6}} \ ,
\label{Ey}
\end{align}
with $n_\rho=0,1,2,\cdots$ and $l=0,1,2,\cdots$.
As discussed above, the fermionic baryons correspond to the
wave functions that are odd in $a_I$, which implies that $l$ should be
odd. We see in the next subsection that this yields baryons with
half-integer spin and isospin.
{}Finally, the quantization of $Z$ is trivial:
\begin{equation}
E_Z=\omega_Z\left(n_z+\half\right)
=\frac{2n_z+1}{\sqrt{6}} \ ,
\label{Ez}
\end{equation}
with $n_z=0,1,2,\cdots$.
Adding Eqs.~\eqref{Ey} and \eqref{Ez}, we obtain the following
baryon mass formula:
\begin{equation}
M=M_0+\sqrt{\frac{(l+1)^2}{6}+\frac{2}{15}N_c^2}
+\frac{2(n_\rho+n_z)+2}{\sqrt{6}}\ .
\label{M1}
\end{equation}

\subsection{Physical interpretation}

The physical interpretation of the baryon spectrum found in the
previous subsection is as follows. As mentioned above, each mass
eigenstate belongs to the $(S_{l/2},S_{l/2})$ representation of
$SO(4)\simeq (SU(2)_I\times SU(2)_J)/\bZ_2$,
which acts on the $SU(2)$-valued collective coordinate $\ba$ defined
by \Eqref{ba} as
\begin{equation}
\ba \to g_I\,\ba\, g_J \ ,\quad
g_{I,J}\in SU(2)_{I,J} \ .
\label{gag}
\end{equation}
This implies that $SU(2)_I$ and $SU(2)_J$ are identified with the
isospin rotation and the spatial rotation, respectively,
as in Ref.~\citen{ANW}. This can be understood from the ansatz
\eqref{def:coll} and \Eqref{aI} in Appendix \ref{metric:inst},
relating $\ba$ and $V$: The spatial rotation of the BPST instanton
configuration \eqref{eq:BPST} gives rise to the transformation of $V$
as
\begin{equation}
V\to Vg_J \ ,\quad g_J\in SU(2)_J\ ,
\end{equation}
while the isospin rotation of the gauge field \eqref{def:coll} is
induced by
\begin{equation}
V\to g_I V\ ,\quad g_I\in SU(2)_I\ .
\end{equation}
This transformation property, together with \Eqref{aI}, implies
\Eqref{gag}. With this identification, we find that the spin $J$ and
isospin $I$ of the soliton are both $l/2$. The $l=1$ states correspond
to $I=J=1/2$ states, which include nucleons, and the $l=3$ states
correspond to $I=J=3/2$ states, which include $\Delta$.
These are the states considered in Ref.~\citen{ANW}.

Heavier baryons with a common spin and isospin are represented by
states with non-trivial $n_\rho$ and $n_z$.
It is interesting that the excited states with odd $n_z$ correspond to
odd parity baryons, as the parity transformation induces $z\to -z$, as
shown in Ref.~\citen{SaSu1}.

For the comparison with our mass formula \eqref{M1} to be made below,
we list baryons with $I=J$ in the PDG baryon summary table \cite{PDG},
along with a possible interpretation of the quantum numbers
$(n_\rho,n_z)$.
{\small
\begin{equation}
\begin{array}{|c||c|cc|cc|cc|ccccc}
\hline
(n_\rho,n_z)&(0,0)&(1,0)&(0,1)&(1,1)&(2,0)/(0,2)&
(2,1)/(0,3)&(1,2)/(3,0)\\
\hline
N\,(l=1)&940^+&1440^+&1535^-&1655^-&1710^+,~?&
2090^-_*,~?&2100^+_*,~?\\
\Delta\,(l=3)&1232^+&1600^+&1700^-&1940^-_*&1920^+,~?
&?,~?&?,~?\\
\hline
\end{array}
\label{PDG}
\end{equation}
}
The superscripts $\pm$ represent the parity.
The subscript $*$ indicates that evidence of the existence of the
baryon in question is poor.

\subsection{Comments on the baryon mass formula}
\label{bmass}

Let us first discuss the $N_c$ dependence of the mass formula
\eqref{M1} in the large $N_c$ limit. For $N_c\gg l$, the mass formula
\eqref{M1} has the following approximate expression:
\begin{equation}
M\simeq M_0+
\sqrt{\frac{2}{15}}N_c
+\frac{1}{4}\sqrt{\frac{5}{6}}\frac{(l+1)^2}{N_c}
+\frac{2(n_\rho+n_z)+2}{\sqrt{6}}\ .
\label{M2}
\end{equation}
Note that the $\cO(N_c)$ terms are identical to the classical formula,
\eqref{Mcl}. It is interesting that the mass formula \eqref{M2} is
consistent with the expected $N_c$ dependence in large $N_c$ QCD
\cite{Witten:1/n,ANW}. It is known that the mass splittings
among the low-lying baryons with different spins are of order
$1/N_c$, while those among excited baryons are of order
$N_c^0$. This is exactly what we observe in \Eqref{M2}.

The states considered in Ref.~\citen{ANW} correspond to the states with
$n_\rho=n_z=0$. The $l$-dependent term in their mass formula is
proportional to $l(l+2)$, which is also reproduced in \Eqref{M2}.

It is important to understand the extent to which we can trust the
mass formulas \eqref{M1} and \eqref{M2}. First, in order to
approximate \Eqref{M1} with \Eqref{M2}, the inequality
\begin{equation}
\frac{(l+1)^2}{6}<\frac{2}{15}N_c^2
\end{equation}
must be satisfied. For real QCD with $N_c=3$, it is satisfied only for
$l=1$. For this reason, we mainly consider the formula \eqref{M1} in
the following. However, we have to keep in mind that there may be
$1/N_c$ corrections to the action \eqref{model} that become important
for large quantum numbers $l$, $n_z$ and $n_\rho$ in the mass formula
\eqref{M1}.

Another uncertainty in the mass formula regards the zero-point
energy. Note that the zero-point energy in \Eqref{M1} is of order
$N_c^0$, which is the same order as possible $1/N_c$
corrections to the classical soliton mass $M_0$.
Furthermore, an infinite number of the heavy modes around the
instanton that have been ignored to this point give a divergent
contribution to the total zero-point energy of order $N_c^0$.
What we really need is the difference between the energy in the
presence of a soliton and that in the vacuum, and hence the
divergence in the zero-point energy in the presence of a soliton
should be removed by subtracting the zero-point energy of the
vacuum. In this paper, we do not attempt to analyze such
contributions. Instead, we only consider the mass differences among
the baryons and treat $M_0$ as a free parameter.

\subsection{Numerical estimates}

As suggested in \S\S \ref{CSterm} and \ref{bmass}, we cannot fully
justify the quantitative prediction for the baryon mass, especially
in the case of large masses, because the contribution from
higher-derivative terms, as well as the $1/N_c$ corrections, may
become important. Nevertheless, here we report some numerical
estimates to gain some insight from the baryon mass formula
\eqref{M1}.

The difference between the masses of the $l=3$ and $l=1$ states is
\begin{equation}
M_{l=3}-M_{l=1}=
\sqrt{\frac{8}{3}+\frac{6}{5}}-\sqrt{\frac{2}{3}+\frac{6}{5}}
\simeq 0.600 \simeq  569\,{\rm MeV}\ .
\end{equation}
The difference between the masses of the $(n_\rho,n_z)=(1,0)$ or
$(0,1)$ state and the $(0,0)$ state with a common $l$ is
\begin{equation}
M_{(1,0)/(0,1)}-M_{(0,0)}=\frac{2}{\sqrt{6}}
\simeq 0.816\simeq 774\,{\rm MeV}\ .
\end{equation}
Here, we have used $1=M_{\rm KK}\simeq 949~{\rm MeV}$, which is
consistent with the $\rho$ meson mass \cite{SaSu1,SaSu2}.
Unfortunately, these values are slightly too large compared with the
experimental values.
If $M_{\rm KK}$ were $500~{\rm MeV}$, the predicted values obtained
using \Eqref{M1} would become very close to those listed in
\eqref{PDG}:\footnote{The nucleon mass $940~{\rm MeV}$ is used as
an input to fix $M_0$.}
{\small
\begin{equation}
\begin{array}{|c||c|cc|cc|cc|ccccc}
\hline
(n_\rho,n_z)&(0,0)&(1,0)&\hspc (0,1)&(1,1)&\hspc (2,0)/(0,2)
&(2,1)/(0,3)&\hspc (1,2)/(3,0)\\
\hline
N\,(l=1)&940^+&1348^+&\hspc 1348^-&1756^-&\hspc 1756^+,1756^+
&2164^-,2164^-&\hspc 2164^+,2164^+\\
\Delta\,(l=3)&1240^+&1648^+&\hspc 1648^-&2056^-&\hspc 2056^+,2056^+
&2464^-,2464^-&\hspc 2464^+,2464^+\\
\hline
\end{array}
\end{equation}
}

\section{Conclusion and discussions}

In this paper, we have investigated dynamical baryons within the
context of the holographic description of QCD proposed in
Refs.~\citen{SaSu1} and \citen{SaSu2}.
A key observation in this treatment is that the baryon number is
provided with the instanton number in the five-dimensional YM-CS
theory \eqref{model}. This implies that baryons can be described as
large $N_c$ solitons, as in Refs.~\citen{Skyrme:1,Skyrme:2,Skyrme:3},
\citen{Witten:1/n} and \citen{ANW}.
We explicitly constructed a soliton solution with a unit baryon
number and found that it corresponds to the BPST instanton with a size
of order $\lambda^{-1/2}$.
It was stressed that the Coulomb interaction in the CS term plays a
crucial role in obtaining the regular solution.
Although regular, the instanton is not large enough that we can employ
the YM-CS theory with all the higher-order derivative terms omitted.
As a first step toward the full incorporation of the infinitely many
higher derivative terms, we consider the non-Abelian DBI action
\cite{Tseytlin} in Appendix \ref{higher}.
There, we verify that the energy contribution of the static baryon
configuration computed with the non-Abelian DBI action is the same as
that computed with the YM action.
We leave the more thorough analysis of this problem as a future
work, with the goal of carrying out a precise quantitative test of the
present model regarding baryon physics by properly treating all the
relevant higher derivative terms. 

We quantized the collective coordinates of the instanton to obtain
the baryon spectrum in the hope that the model \eqref{model} captures
some qualitative features of baryons.
In fact, the $N_c$ dependence of the baryon mass formula \eqref{M1}
is consistent with the results of the analyses of large $N_c$ baryons
in the literature. Furthermore, our model describes negative-parity
baryons as the excited states of the instanton along the holographic
direction $z$. Unfortunately, the best fit of the parameter $\Mkk$ to
the experimental data for baryons is inconsistent with that found in
Refs.~\citen{SaSu1} and \citen{SaSu2}, which comes from the $\rho$
meson mass. This may be due to the fact that the higher derivative
terms have not been incorporated into the YM-CS theory.

We end this paper with some comments on future directions.
It is important to analyze static properties of baryons, such as the
charge radii and magnetic moments, as done in Ref.~\citen{ANW} for the
Skyrme model.
Also, extension of the one-instanton solution to multi-instanton cases
is quite interesting for the purpose of exploring multi-baryon
systems. (See Refs.~\citen{AtMa} and \citen{Hosaka:1990dg} for related
works.)
Moreover, in the present model, the role of the infinite number of
(axial-)vector mesons in obtaining the soliton solution is not
difficult to elucidate. It would be interesting to compare this role
with the recent analysis given in Ref.~\citen{NaSuKo}, in which
baryons are constructed as Skyrmions in the effective action including
the pion and $\rho$ meson on the basis of the D4/D8 model.

\section*{Acknowledgements}

We would like to thank our colleagues in the particle theory groups at
Kyoto University, Nagoya University and Ibaraki University for
discussions and encouragement.
S.~S.\ and S.~Y.\ are also grateful to H.~Suganuma, K.~Nawa and
T.~Kojo for useful discussions during the workshop YKIS2006,
``New Frontiers in QCD".
This work is supported by a Grant-in-Aid for the 21st Century
COE ``Center for Diversity and Universality in Physics", from the
Ministry of Education, Culture, Sports, Science and Technology
(MEXT) of Japan.
T.~S.\ and S.~S.\ would like to thank the Yukawa Institute for
Theoretical Physics, Kyoto University, where part of this
work was done, for kind hospitality.
Numerical analyses in the early stages of this work were carried out
on Altix3700 BX2 at YITP, Kyoto University.
The work of H.~H.\ was supported in part by a Grant-in-Aid for
Scientific Research (C) No.\ 18540266 from the Japan Society for the
Promotion of Science (JSPS).
The work of T.~S.\ and S.~S.\ was partly supported by Grants-in-Aid
for Young Scientists (B) Nos.\ 18740126 and 17740143, respectively,
from the Ministry of Education, Culture, Sports, Science and
Technology, Japan.

\appendix

\section{Metric of the Instanton Moduli Space}
\label{metric:inst}

In this appendix, we outline the derivation of the metric of the
instanton moduli space given by \Eqref{metric}. (See, e.g.,
Ref.~\citen{instanton:review} for a review.)
This metric can be read from the kinetic term of the Lagrangian of
the collective coordinates, which follows from the $F_{0M}^2$ term in
\Eqref{DBI} as
\begin{align}
\frac{m_X}{2}\, g_{\alpha\beta}\dot{X}^\alpha\dot{X}^\beta
&=\kappa\lambda^{-1}\int d^3xdz\,\tr F_{0M}^2\nn\\
&=\kappa\lambda^{-1}\int d^3xdz\tr
\left(D_M^{\rm cl}\Phi-\dot{A}_M^{\rm cl}\right)^2\ ,
\label{eq:calcmetric}
\end{align}
where $F_{0M}$ is given by \Eqref{F0M}.
We solve the Gauss's law constraint \eqref{Gausslaw2} to obtain $\Phi$
(\Eqref{VV}) for each instanton moduli and then calculate the
corresponding metric using \Eqref{eq:calcmetric}.
To do this, we first decompose $\Phi$ as
\begin{equation}
\Phi=\Phi_X+\Phi_\rho+\Phi_{SU(2)}\ ,
\end{equation}
and impose the conditions
\begin{align}
&D_M^{\rm cl}\left(\dot X^N
\frac{\del}{\del X^N} A_M^{\rm cl}
-D_M^{\rm cl}\Phi_X\right)=0 \ ,
\label{Gausslaw3-1}\\
&D_M^{\rm cl}\left(
\dot\rho \frac{\del}{\del \rho} A_M^{\rm cl}
-D_M^{\rm cl}\Phi_\rho\right)=0 \ ,
\label{Gausslaw3-2}\\
&D_M^{\rm cl}D_M^{\rm cl}\Phi_{SU(2)}=0 \ .
\label{Gausslaw3-3}
\end{align}
The following formula for $g(x)$ given in \Eqref{eq:g} is useful in
the derivation below:
\begin{equation}
g\del_M g^{-1}=
\begin{cases}
\ds \frac{i}{\xi^2}\left((z-Z)\tau^i
-\epsilon_{ija}(x^j-X^j)\tau^a\right)\ ,
& (M=i)\\
\ds -\frac{i}{\xi^2}(x^a-X^a)\tau^a \ .
& (M=z)
\end{cases}
\end{equation}

\noindent
$\bullet$ Instanton center $X^M=(\vec{X},Z)$

We find that the form
\begin{equation}
\Phi_X=-\dot X^NA_N^{\rm cl}
\end{equation}
satisfies \Eqref{Gausslaw3-1}, since we have
$(\del/\del X^N)A_M^{\rm cl}=-\del_NA_M^{\rm cl}$,
and hence
\begin{equation}
D_M^{\rm cl}\Phi_X-\dot X^N\frac{\del}{\del X^N}A_M^{\rm cl}
=-\dot X^N F_{MN}^{\rm cl} \ .
\end{equation}
The corresponding metric is
\begin{equation}
g_{MN}=\frac{2\kappa\lambda^{-1}}{m_X}\int\! d^3x\,dz
\tr F^{\rm cl}_{MP}F^{\rm cl}_{NP}=\delta_{MN}\ .
\end{equation}

\noindent
$\bullet$ Instanton size $\rho$

Using the relation
\begin{equation}
\frac{\del}{\del \rho} A_M^{\cl}
=-\frac{2\rho}{\xi^2+\rho^2}A_M^{\cl}
\end{equation}
and the formula
\begin{equation}
\del_M(g\,\del_M\,g^{-1})
\propto(x_M-X_M)\,g\,\del_M\,g^{-1}=0 \ ,
\end{equation}
we find that \Eqref{Gausslaw3-2} is satisfied by
\begin{equation}
\Phi_\rho=0\ .
\end{equation}
Further, the metric is given by
\begin{equation}
g_{\rho\rho}=\frac{2\kappa\lambda^{-1}}{m_X}
\int\! d^3x\,dz
\tr\left(\frac{\del}{\del\rho}A^{\rm cl}_M\right)^2
=2\ .
\end{equation}

\noindent
$\bullet$ $SU(2)$ orientation

The $SU(2)$ rotation of the instanton solution is implemented
by a global gauge transformation. To solve \Eqref{Gausslaw3-3}, it is
convenient to move to the singular gauge obtained through the gauge
transformation
\begin{align}
\Phi_{SU(2)}&\to \Phib_{SU(2)}\equiv g^{-1}\Phi_{SU(2)}\, g\ ,
\nn\\
A_M^{\rm cl}&\to \Ab_M\equiv
g^{-1}A_M^{\rm cl}\, g -ig^{-1}\del_M g
=-i(1-f(\xi))g^{-1}\del_M g\ ,
\end{align}
where $f(\xi)$ is given by \Eqref{eq:f}.
Then, \Eqref{Gausslaw3-3} can be recast as
\begin{equation}
\Db_M\Db_M\Phib_{SU(2)}=0 \ ,
\label{eq:DbDbPhi_a=0}
\end{equation}
with $\Db_M=\del_M+i[\Ab_M \ ,\ ]$.
It is not difficult to see that \Eqref{eq:DbDbPhi_a=0} is solved by
\begin{equation}
\Phib_a=u(\xi)\frac{\tau^a}{2}\ ,
\quad (a=1,2,3)
\end{equation}
with $u(\xi)$ satisfying
\begin{equation}
\frac{1}{\xi^3}\del_{\xi}(\xi^3\del_{\xi}\, u(\xi))
=8\frac{(1-f(\xi))^2}{\xi^2}u(\xi) \ .
\end{equation}
The regular solution of this equation is
\begin{equation}
u(\xi)=C\,\frac{\xi^2}{\xi^2+\rho^2}=C\, f(\xi)\ ,
\end{equation}
with a constant $C$. Therefore, $\Phi_{SU(2)}$ can be written
\begin{equation}
\Phi_{SU(2)}=\chi^a(t)\Phi_a(x)\ ,
\label{chiphi}
\end{equation}
with
\begin{equation}
\Phi_a=f(\xi)\,g\,\frac{\tau^a}{2}\,g^{-1}
\label{phia}
\end{equation}
and $t$-dependent real coefficients $\chi^a(t)$.

We choose the $SU(2)$-valued collective coordinate
$\ba(t)=a_4(t)+ia_a(t)\tau^a$ as
\begin{equation}
V(t,\vec x,z)\to\ba(t)\ .\quad(z\to\infty)
\label{aI}
\end{equation}
Comparing this with \eqref{chiphi}, we find
\begin{equation}
\chi^a=-i\tr\left(\tau^a\ba^{-1}\dot \ba\right)
=2\left(a_4\dot a_a-\dot a_4 a_a+\epsilon_{abc}a_b\dot a_c
\right)\ .
\label{chia}
\end{equation}
This gives
\begin{equation}
\left(\chi^a\right)^2=4\dot{a}_I^2 \ .
\end{equation}
Then, the metric for $a_I$ is obtained as
\begin{equation}
g_{IJ}\dot a_I\dot a_J
=\frac{2\kappa\lambda^{-1}}{m_X}\int d^3xdz
\tr\left(D^{\rm cl}_M\Phi_{SU(2)}\right)^2=2\rho^2\dot a_I^2\ ,
\end{equation}
with the constraint $a_I^2=1$.

It is easy to see that the off-diagonal components of
$g_{\alpha\beta}$, connecting different kinds of moduli, vanish.
Collecting these results, we find that the metric of the moduli space
is given by
\begin{align}
ds^2&=g_{\alpha\beta}dX^\alpha dX^\beta\nn\\
&=d\vec X^2+dZ^2+2(d\rho^2+\rho^2 da_I^2)\nn\\
&=d\vec X^2+dZ^2+2\, dy_I^2\ ,
\end{align}
with $y_I=\rho a_I$.
Note that $a_I$ parameterizes not $S^3$ but $S^3/\bZ_2$, with
$\bZ_2$ being the center of $SU(2)$, which acts as $a_I\to -a_I$.
In fact, the configuration \eqref{def:coll} is unchanged under
the $\bZ_2$ transformation $V\to -V$. Hence, the one-instanton moduli
space coincides with $\bR^4\times\bR^4/\bZ_2$.

\section{Higher Derivative Terms}
\label{higher}

As we have seen in \S \ref{solution}, higher derivative
terms in the D-brane action also contribute to the soliton mass
at the same order in the $1/\lambda$ expansion.
However, it is difficult to include all the higher-order derivative
terms, since the exact derivative corrections in the D-brane action
are not known. (See, e.g., Ref.~\cite{higher_derivative_terms} and the
references therein.)
Among the various sources of higher derivative corrections in the
D-brane action, here we consider the contributions from the
non-Abelian DBI action \cite{Tseytlin} as a first step toward a
complete analysis.

The non-Abelian DBI action for the probe D8-branes is given by
\begin{equation}
S_{\rm {DBI}}=-\mu_8\int d^9x\,e^{-\phi}\,{\rm str}
\sqrt{-\det(g_{ab}+2\pi\alpha'\cF_{ab})} \ ,
\quad(a,b=0,1,\cdots,8)
\label{NBI}
\end{equation}
where $\mu_8=1/((2\pi)^8l_s^9)$, and ${\rm str}$ denotes the
symmetrized trace. Here, $g_{ab}$ is the induced metric on the
D8-brane world-volume, given by \cite{SaSu1}
\begin{equation}
ds^2_{\rm 9\,dim}=\frac{\lambda l_s^2}{3}\left[\,
\frac{4}{9}k(z)^{1/2}\,\eta_{\mu\nu}dx^\mu dx^\nu
+\frac{4}{9}k(z)^{-5/6}\,dz^2+k(z)^{1/6}\,d\Omega_4^2\,\right]\ ,
\end{equation}
and the dilaton on it reads
\begin{equation}
e^{-\phi}=\frac{3^{3/2}\pi N_c}{\lambda^{3/2}}
k(z)^{-1/4}\ .
\end{equation}
After integrating over the $S^4$ directions, we obtain
the five-dimensional non-Abelian DBI action
\begin{equation}
S_{\rm DBI}=
-\frac{N_c\lambda^3}{3^9\pi^5}
\int d^4xdz \,k(z)^{1/12}\,{\rm str}
\sqrt{-\det\left(g^{(5)}_{\hat M\hat N}
+\frac{27\pi}{2\lambda}\cF_{\hat M\hat N}\right)}\ ,
\label{5dNBI}
\end{equation}
where $\hat M,\hat N=0,1,2,3,z$, and the five-dimensional
metric $g^{(5)}_{\hat M\hat N}$ is given by
\begin{equation}
ds^2_{\rm 5\,dim}=k(z)^{1/2}\,\eta_{\mu\nu}dx^\mu dx^\nu
+k(z)^{-5/6}\, dz^2\ .
\end{equation}
Here, for simplicity, we have kept only the gauge potentials
$A_{\hat M}$ non-zero.

The above expressions are written in terms of the original variables,
before the rescaling \eqref{rescaling}.
Upon the rescaling, it is found that the non-Abelian DBI action
\eqref{5dNBI} can be expanded as
\begin{align}
S_{\rm DBI}
&=-\frac{\lambda N_c}{3^9\pi^5}\int d^4xdz
\left(\cL_0+\lambda^{-1}\cL_1
+\cO(\lambda^{-2})\right) \ ,
\end{align}
where $\cL_0$ and $\cL_1$ are given by
\begin{align}
\cL_0&={\rm str}\sqrt{\det\left(\cB_{MN}\right)} \ ,
\label{L0}
\\
\cL_1&={\rm str}\Biggl[
\sqrt{\det\left(\cB_{MN}\right)}
\Biggl(
\frac{2}{3}z^2-\half\left(\frac{27\pi}{2}\right)^2
\cG^{MN}\cF_{M0}\cF_{N0}
\nn\\
&\qquad\quad
+\frac{z^2}{2}\left(\frac{27\pi}{2}\right)^2
\left\{-\half\cG^{MN}\cF_{Mi}\cF_{Ni}+
\frac{5}{6}\cG^{MN}\cF_{Mz}\cF_{Nz}
\right\}\Biggr)
\Biggr] \ ,
\end{align}
with $M,N=1,2,3,z$ and $i=1,2,3$, and we have the definitions
\begin{align}
\cB_{MN}\equiv\delta_{MN}+\frac{27\pi}{2}\cF_{MN} \ ,\quad
\cG^{MN}\equiv (\cB^{-1})^{(MN)}\ .
\end{align}
{}From \Eqref{L0}, we find that the leading-order term in the
$1/\lambda$ expansion is given by the non-Abelian DBI action in a flat
space-time. It is known that the BPST instanton configuration
\eqref{eq:BPST} is a solution for the non-Abelian DBI action
\eqref{L0} \cite{9711097,9801127,9804180}.

Inserting the BPST instanton configuration \eqref{eq:BPST}
into the action \eqref{5dNBI}, we obtain
\begin{align}
S_{\rm DBI}=&-\frac{\lambda N_c}{3^9\pi^5}\int\!d^4x\,dz\,
k^{2/3}\nn\\
&\quad\times2\left(1+2\frac{\del}{\del s}\right)
\sqrt{1+s\left(\frac{27\pi}{4}\right)^2k^{-1}\omega^2}
\sqrt{1+s\left(\frac{27\pi}{4}\right)^2k^{1/3}\omega^2}
\,\Bigg|_{s=1}\ ,
\end{align}
where
\begin{equation}
\omega=\frac{4\rho^2}{(\xi^2+\rho^2)^2}\ ,
\end{equation}
and $\xi$ is defined as in \Eqref{eq:f}.
Here, we are using the rescaled variables, and we have
$k=k(\lambda^{-1/2}z)=1+\lambda^{-1}z^2$.
Hence, the energy contribution is
\begin{align}
E&=\frac{\lambda N_c}{3^9\pi^5}\int\!d^3x\,dz\,k^{2/3}
\nn\\
&\qquad\left.\times2\left(1+2\frac{\del}{\del s}\right)\left(
\sqrt{1+s\left(\frac{27\pi}{4}\right)^2k^{-1}\omega^2}
\sqrt{1+s\left(\frac{27\pi}{4}\right)^2k^{1/3}\omega^2}
-1\right)\,\right|_{s=1}
\nn\\
&=\frac{\lambda N_c}{18\pi^2}\int_{-\infty}^\infty\!dz
\int_0^\infty\!dr\,
\left[r^2\left(1+\frac{z^2}{3\lambda}\right)\omega^2
+\cO(\lambda^{-2})\right]
\nn\\
&=8\pi^2\kappa\left[1+
\lambda^{-1}\left(
\frac{\rho^2}{6}+\frac{Z^2}{3}\right)
+\cO(\lambda^{-2})\right]\ ,
\label{rhoZ}
\end{align}
where we have used
\begin{equation}
\int_{-\infty}^\infty\!dz\int_0^\infty\!dr \,r^2\omega^2
=\frac{2\pi}{3}\ ,
\quad\int_{-\infty}^\infty\!dz\int_0^\infty\!dr \,r^2z^2\omega^2=
\frac{2\pi}{3}Z^2+\frac{\pi}{3}\rho^2\ .
\end{equation}
The expression \eqref{rhoZ} is identical to \eqref{smass}, except for
the contribution from the CS term. This result is highly non-trivial,
since the non-Abelian DBI action \eqref{5dNBI} contains infinitely
many higher derivative terms, while our previous analysis is based on
the YM action in \Eqref{model}.
This finding suggests that our previous results may not be
significantly modified even if we include all of the higher derivative
terms. However, because there are still infinitely many higher derivative
terms that have not been included in the non-Abelian DBI action, we
cannot definitively confirm the quantitative results obtained in this
paper, such as the baryon mass formula \eqref{smass}. Nonetheless,
we expect that these results will be useful in more systematic studies
of the higher derivative terms.

\vspace{5mm}
{\footnotesize
\noindent
{\bf Note added:}
While this paper was being completed, we received the paper
\cite{HRYY}, whose content overlaps somewhat with that of the present
paper.
}

\end{document}